\documentclass[preprint,prb]{revtex4}
\usepackage{graphicx}
\usepackage{latexsym}
\usepackage{amsfonts}
\usepackage{revsymb}
\begin{document}

\title{A Confining Non-Local Four-Fermi Interaction from Yang-Mills Theory in a Stochastic Background}

\author{Jose A. Magpantay}
\email{jose.magpantay@upd.edu.ph}

\affiliation{ National Institute of Physics, University of the
Philippines, Diliman Quezon City, 1101,Philippines }

\date{\today}

\begin{abstract}
We derive a non-local four-fermi term with a linear potential from
Yang-Mills theory in a stochastic background.  The stochastic
background is a class of classical configuration derived from the
non-linear gauge.
\end{abstract}

\maketitle

\section{Introduction}

The structure of the vacuum plays a key role in understanding
physical phenomena.  For example, QED and the asymptotic freedom
phase of non-Abelian gauge theory are based on the trivial vacuum
$(A_{\mu} = 0)$ and the physical vector fields, the transverse
photon/gluon, are clearly exposed by using the Coulomb gauge.
 Confinement, on the other hand, is believed to be a
 non-perturbative vacuum phenomenon that is not yet clearly
 understood.  Presently, there are two confinement mechanisms
 discussed extensively in the literature.  These are the magnetic
 monopoles\cite{1,2,3,4} and vortices\cite{5,6}, which are related to the
 maximal Abelian gauge\cite{7} and maximal central gauge\cite{8},
 respectively.  Although each of the two mechanism has its own set of successes
 and failures, they share an important shortcoming: they
 are disjointed from short-distance physics.  As a quark is pulled
 from inside a hadron, it goes through the asymptotic phase
 (short-distance, where transverse gluons are exchanged) to the
 confinement phase (large-distance, with mass gap and linearly rising interaction)
 eventually resulting in hadronization.  Ideally then, the asymptotic phase and the
 confinement phase should continuously interpolate  in a way where
 the effective degrees of freedom in each distance regimes are
 transparent.  On the lattice, this problem is dealt with using a decimation
 procedure.\cite{9}
 For a number of years now, this author has been
 claiming that this is achieved in the continuum by using the non-linear gauge.

 The non-linear gauge\cite{10}, which is given by
\begin{equation}\label{eq:1}
(\partial\cdot D)^{ab}(\partial\cdot
A^{b})=(D\cdot\partial)^{ab}(\partial\cdot
A^{b})=(\partial^{2}\delta^{ab}-g\epsilon^{abc}A^{c}\cdot\partial)(\partial\cdot
A^{b})=0,
\end{equation}
was proposed by the author eleven years ago because of the
observation that field configurations that satisfy $\partial\cdot
A^{a}=f^{a}(x)\neq 0$ and equation (\ref{eq:1}) cannot be
gauge-transformed to the Coulomb gauge.  This follows from the
fact that the zero mode of $(\partial\cdot D)^{ab}$, which is
$\partial\cdot A^{a}$, is also the source in the equation that
solves for the gauge parameter.  And conversely, field
configurations on the Coulomb surface cannot be gauge transformed
to the non-linear regime of equation (\ref{eq:1}), i.e., there is
no $\Lambda^{a}$ such that $A^{'}_{\mu}=A_{\mu}+D_{\mu}\Lambda$,
$\partial\cdot A^{'}\neq 0$, and $(\partial\cdot
D(A^{'})(\partial\cdot A^{'})=0$.  These mean that the two regimes
of equation (\ref{eq:1}), which are (i) the linear regime with
$\partial\cdot A^{a}=0$, and (ii) the non-linear regime, which
effectively is the Gribov horizon of the surface $\partial\cdot
A^{a}=f^{a}(x)$, do not mix.  Thus, the Coulomb gauge is an
incomplete gauge-fixing for non-Abelian theory, although there are
claims to the contrary.\cite{11}

Using the running of the coupling constant, the author argued that
the non-linear gauge interpolates between short-distance
(perturbative) physics and large-distance (non-perturbative)
physics.  In short-distance regime, where the Coulomb gauge is in
effect, the transverse gluons are the physical degrees of freedom.
However, in the large-distance regime, the vector field
$t^{a}_{\mu}$ and the scalar field $f^{a}$, which satisfy two sets
of constraints, are the effective degrees of freedom.\cite{12} The
full quantum dynamics of the scalars lead to a Parisi-Sourlas
mechanism\cite{13} with an $O(1,3)$ symmetry\cite{14}, which
unfortunately has a wrong sign for the kinetic term of the
scalars.  Thus, although we have shown equivalence to a 2D O(1,3)
non-linear sigma model, the proof of confinement remains formal
because aside from the wrong sign of the scalar kinetic term, the
mechanism is not identified and quarks and "gluons" are not
involved.

To identify the mechanism, the author proposes that we look for
(a) classical solution(s) of the pure $f^{a}$ dynamics which is
(are) vacuum configuration(s). We then expand on this (these)
configuration(s) and see the physics of quarks, "gluons" and
scalar fluctuations around the background. This program was
started by the author about five years ago and the following
results had been derived.  First, that all spherically symmetric
functions $\tilde{f}^{a}(x)$, with $x=(x_{\mu}x_{\mu})^{1/2}$, in
$R^{4}$ are classical solutions with zero field strength.\cite{15}
This shows that all the $\tilde{f}^{a}(x)$ are vacuum
configurations with zero action, which means the configuration
space has a very broad minimum.

Second, since we do not have just one classical configuration but
a whole class of solutions, the author proposed to treat them as a
stochastic background with a white-noise distribution.  This led
to an area law behaviour for the Wilson loop.\cite{15}  Thus, we
have identified a possible mechanism for confinement.

Third, when we consider the classical dynamics of the "gluons" in
the large distance regime, we find the gluons acquiring a
mass.\cite{16} This verifies the existence of a mass gap.
Furthermore, the "gluons" also lose their self-interaction.  This
result was also arrived at by Kondo by a suitable redefinition of
fields.\cite{17}

In a forthcoming paper\cite{18}, which will provide the details of
the results presented here and discuss quantum field theory in a
stochastic background, the author will show that the scalars
$\phi^{a}(x)$, which are fluctuations off the classical stochastic
background $\tilde{f}^{a}(x)$,
\begin{equation}\label{eq:2}
f^{a}=\tilde{f}^{a}(x)+\phi^{a}(x)
\end{equation}
are non-propagating, i.e., they do not have a kinetic term.  As a
matter of fact, we will show that $\langle
S_{YM}(A=A(\tilde{f}+\phi, t^{a}_{\mu}))\rangle_{\tilde{f}}$ is
independent of the scalars $\phi^{a}$ and only dependent on
$t^{a}_{\mu}$.  This resolves the problem of scalars with a wrong
sign for the kinetic term encountered in the derivation of the
Parisi-Sourlas mechanism, where the full quantum theory of the
scalars was shown to be equivalent to a 2D O(1,3) non-linear
$\sigma$ model.

In this paper, we will present the quantum dynamics of quarks and
"gluons" in a stochastic background, resulting mainly in the
derivation of a non-local four-fermi interaction with a linear
potential.

\section{Quarks and "Gluons" in the Non-Linear Regime}

In this section, we will provide the background needed to derive
the main result which will be discussed in section 3.

Consider $SU(2)$ theory with action
\begin{equation}\label{eq:3}
S=S_{YM}+S_{fermion}=\int
d^{4}x[\frac{1}{4}F^{a}_{\mu\nu}F^{a}_{\mu\nu}+\Psi i
\gamma_{\mu}D_{\mu}\psi]=\int\mathcal{L}d^{4}x
\end{equation}
 The gauge-fixing given by equation(\ref{eq:1}) leads to the
resolution of the potential in the non-linear regime as given by
\begin{equation}\label{eq:4}
A^{a}_{\mu}(x)=\frac{1}{(1+\vec{f}\cdot\vec{f})}(\delta^{ab}+\epsilon^{abc}f^{c}+f^{a}f^{b})(\frac{1}{g}\partial_{\mu}f^{b}+t^{b}_{\mu})
\end{equation}
The scalars $f^{a}$ and the "gluon" $t^{a}_{\mu}$ satisfy the
following constraints, which guarantee the same number of degrees
of freedom,
\begin{eqnarray}\label{eq:5}
\partial\cdot t^{a}&-&\frac{1}{g\ell^{2}}f^{a}=0,\\
\rho^{a}&=&\frac{1}{(1+\vec{f}\cdot\vec{f})^{2}}[\epsilon^{abc}+\epsilon^{abd}f^{d}f^{c}-\epsilon^{acd}f^{d}f^{b}+f^{a}f^{d}\epsilon^{dbc}\nonumber\\
&-&f^{a}(1+\vec{f}\cdot\vec{f})\delta^{bc}-f^{c}(1+\vec{f}\cdot\vec{f})\delta^{ab}]\partial_{\mu}f^{b}t^{c}_{\mu}=0.\label{eq:6}
\end{eqnarray}
Substituting equation (\ref{eq:4}) in the field strength, we find
\begin{eqnarray}
F^{a}_{\mu\nu}&=&\frac{1}{g}Z^{a}_{\mu\nu}(f)+L^{a}_{\mu\nu}(f;t)+gQ^{a}_{\mu\nu}(f;t),\label{eq:7}\\
Z^{a}_{\mu\nu}&=&X^{abc}\partial_{\mu}f^{b}\partial_{\nu}f^{c},\label{eq:8}\\
L^{a}_{\mu\nu}(f,t)&=&R^{ab}(\partial_{\mu}t^{b}_{\nu}-\partial_{\nu}t^{b}_{\mu})+Y^{abc}(\partial_{\mu}f^{b}t^{c}_{\nu}-\partial_{\nu}f^{b}t^{c}_{\mu}),\label{eq:9}\\
Q^{a}_{\mu\nu}(f;t)&=&T^{abc}t^{b}_{\mu}t^{c}_{\nu},\label{eq:10}\\
X^{abc}&=&\frac{1}{(1+\vec{f}\cdot\vec{f})^{2}}[-(1+2\vec{f}\cdot\vec{f})\epsilon^{abc}+2\delta^{ab}f^{c}-2\delta^{ac}f^{b}\nonumber\\
&+&3\epsilon^{abd}f^{d}f^{c}-3\epsilon^{acd}f^{d}f^{b}+\epsilon^{bcd}f^{a}f^{d}],\label{eq:11}\\
R^{ab}&=&\frac{1}{(1+\vec{f}\cdot\vec{f})}(\delta^{ab}+\epsilon^{abc}f^{c}+f^{a}f^{b}),\label{eq:12}\\
Y^{abc}&=&\frac{1}{(1+\vec{f}\cdot\vec{f})^{2}}[-(\vec{f}\cdot\vec{f})\epsilon^{abc}+(1+\vec{f}\cdot\vec{f})f^{a}\delta^{bc}-(1-\vec{f}\cdot\vec{f})\delta^{ac}f^{b}\nonumber\\
&+&3\epsilon^{cad}f^{d}f^{b}-2f^{a}f^{b}f^{c}+\epsilon^{abd}f^{d}f^{c}+f^{a}\epsilon^{bcd}f^{d}],\label{eq:13}\\
T^{abc}&=&\frac{1}{(1+\vec{f}\cdot\vec{f})}[\epsilon^{abc}+(1+\vec{f}\cdot\vec{f})f^{b}\delta^{ac}-(1+\vec{f}\cdot\vec{f})\delta^{ab}f^{c}\nonumber\\
&+&\epsilon^{abd}f^{d}f^{c}+f^{a}\epsilon^{bcd}f^{d}+\epsilon^{adc}f^{d}f^{b}]\label{eq:14}.
\end{eqnarray}

The pure $f^{a}$ dynamics given by the action
\begin{equation}\label{eq:15}
S_{1}(f)=\frac{1}{4g^{2}}\int d^{4}x Z^{a}_{\mu\nu}Z^{a}_{\mu\nu},
\end{equation}
hints of non-perturbative physics because (i) of the $1/g^{2}$
factor that goes with $S_{1}$, (ii) the action is infinitely
non-linear, and (iii) $S_{1}\sim (\partial f)^{4}$.  For these
reasons, we look for a classical solution to
\begin{equation}\label{eq:16}
\frac{\delta S_{1}}{\delta f^{a}}=\frac{\delta
Z^{b}_{\mu\nu}}{\delta f^{a}}Z^{b}_{\mu\nu}=0.
\end{equation}
From equation (\ref{eq:8}) and (\ref{eq:11}), we find that all
spherically $\tilde{f}^{a}(x)$, with $x=(x_{\mu}x_{\mu})^{1/2}$
are classical solutions with zero field strength $Z^{a}_{\mu\nu}$.
This follows from
$\partial_{\mu}\tilde{f}^{a}=\frac{x_{\mu}}{x}\frac{d\tilde{f}^{a}}{dx}$
and the anti-symmetry of $X^{abc}$ with respect to b and c. Thus,
we are dealing with a whole class of classical solutions with zero
field strength. This shows that the action has a very broad bottom
(note $S_{1}$ is positive semi-definite).

Since we do not have just one solution but a whole class of
solutions, it was proposed in reference (7) to treat
$\tilde{f}^{a}(x)$ as a stochastic background with a white-noise
distribution
\begin{equation}\label{eq:17}
P[\tilde{f}]=\mathcal{N}exp.(-\frac{1}{\ell}\int^{\infty}_{0}\tilde{f}^{a}(s)\tilde{f}^{a}(s)ds).
\end{equation}
We identify $\ell$ as the length scale where non-perturbative
physics becomes important.  From the running of the coupling
constant, we deduce $\ell=\Lambda^{-1}_{QCD}$.  Substituting the
background decomposition given by equation (\ref{eq:2}) in
$S_{YM}$, we will get a rather involved expression for the action
of the "gluon" $t^{a}_{\mu}$ and the scalars $\phi^{a}$ in the
presence of the background $\tilde{f}^{a}$.  However, as we show
in reference (18), all the stochastic averages drop out except the
following simple result
\begin{equation}\label{eq:18}
\langle S_{YM}\rangle_{\tilde{f}}=\frac{1}{4}\int
d^{4}x\{\frac{1}{3}(\partial_{\mu}t^{a}_{\nu}-\partial_{\nu}t^{a}_{\mu})^{2}+\frac{3}{2}(\frac{n}{\ell})^{2}t^{a}_{\mu}t^{a}_{\mu}\}.
\end{equation}
This result shows that the scalars $\phi^{a}$ are non-propagating
and the "gluons" $t^{a}_{\mu}$ acquired a mass and lost their
self-interactions.

The stochastic average of the fermionic term, on the other hand,
yields
\begin{equation}\label{eq:19}
\langle S_{fermion}\rangle_{\tilde{f}}=\int
d^{4}x\bar{\psi}i\gamma_{\mu}[\partial_{\mu}-gT^{a}(\frac{1}{3}t^{a}_{\mu}+\frac{1}{3g}\partial_{\mu}\phi^{a}
-\frac{2}{3}g(\frac{n}{\ell})\frac{x_{\mu}}{x}\phi^{a})]\psi(x).
\end{equation}

We can remove the scalars from equation (\ref{eq:19}) by using the
stochastic average of the constraints given by equations
(\ref{eq:5}) and (\ref{eq:6}), which are
\begin{eqnarray}
\partial\cdot t^{a}&=&\frac{1}{g\ell^{2}}\phi^{a},\label{eq:20}\\
x\cdot t^{a}&=& 0\label{eq:21}
\end{eqnarray}

Substituting equation (\ref{eq:20}) in equation (\ref{eq:19}), we
find that stochastically averaged effective dynamics yield the
following for fermi interaction
\begin{eqnarray}\label{eq:22}
S_{FF}=g^{2}\int d^{4}x
d^{4}y(\bar{\psi}i\gamma_{\mu}T^{a}\psi)_{x}(\frac{1}{3}\delta_{\mu\alpha}+\frac{1}{3}\ell^{2}\overrightarrow{\partial}_{\mu}\overrightarrow{\partial}_{\alpha}-\frac{2}{3}g^{2}
(\frac{n}{\ell})^{2}\frac{x_{\mu}}{x}\overrightarrow{\partial}_{\alpha})_{x}\nonumber\\
\langle
0|T(t^{a}_{\alpha}(x)t^{a'}_{\beta}(y))|0\rangle(\frac{1}{3}\delta_{\nu\beta}+\frac{1}{3}\ell^{2}\overleftarrow{\partial}_{\nu}\overleftarrow{\partial}_{\beta}-\frac{2}{3}g^{2}(\frac{n}{\ell})^{2}\frac{y_{\nu}}{y}\overleftarrow{\partial}_{\beta})
(\bar{\psi}i\gamma_{\nu}T^{a'}\psi)_{y}
\end{eqnarray}
The Greens function $G_{\alpha\beta}(x-y)$, which is the
propagator of the "gluons" as given by
\begin{eqnarray}\label{eq:23}
\delta^{aa'}G_{\alpha\beta}(x-y)&=&\langle 0
|T|t^{a}_{\alpha}(x)t^{a'}_{\beta}(y)|0\rangle\nonumber\\
&=&\delta^{aa'}\delta_{\alpha\beta}\tilde{G}(x-y)-\frac{1}{m^{2}}\delta^{aa'}\partial^{x}_{\alpha}\partial^{x}_{\beta}\tilde{G}(x-y),
\end{eqnarray}

where $m=\frac{n}{\ell}$ and $\tilde{G}(x-y)$
satisfies\footnote{In reference~\cite{16}, the negative sign from
integration by parts was omitted resulting in the wrong sign
$(+\Box^{2})$ instead of $-\Box^{2}$ in field equation.
Furthermore, since the scalar fluctuations $\phi^{a}$ where
neglected, the gluons satisfied the constraint
$\partial_{\mu}t^{a}_{\mu}=0$, instead of equation~\ref{eq:20},
thus resulting in an inconsistent massive but transverse gluons in
reference~\cite{16}.  This is now corrected in this paper.}
\begin{equation}\label{eq:24}
(-\square^{2}+m^{2})\tilde{G}(x-y)=\delta^{4}(x-y).
\end{equation}

The solution to equation (\ref{eq:24}) for large $|x-y|$
is\cite{20}
\begin{equation}\label{eq:25}
\tilde{G}(x-y)\sim\frac{1}{|x-y|^\frac{3}{2}}e^{-m|x-y|}.
\end{equation}
Substituting equations (\ref{eq:25}), (\ref{eq:23}) in equation
(\ref{eq:22}), we find that the effective four-fermi term from the
"gluon" exchange does not confine because the exponential decay
behaviour dominates. The confining non-local four-fermi term must
come from elsewhere, which must involve the vacuum configuration
at the outset.

\section{Derivation of the Confining Four-Fermi Interaction}

The path-integral in the non-linear regime of the non-linear gauge
is given by\cite{14}
\begin{eqnarray}\label{eq:26}
W=\int(dt^{a}_{\mu})(df^{a})(d\psi)(d\bar{\psi})\delta(\partial\cdot
t^{a}-\frac{1}{g\ell^{2}}f^{a})\delta(\rho^{a})\nonumber\\
\times det^{4}(1+\vec{f}\cdot\vec{f})det\theta
exp.\{-(S_{YM}+S_{fermion})\}
\end{eqnarray}
where $\rho^{a}$ is given by equation (\ref{eq:6}) and $\Theta$ is
the dimension 4 "Fadeev-Popov" operator\cite{15} given by
\begin{equation}\label{eq:27}
\theta^{ad}=(D\cdot\partial)^{ab}(\partial\cdot
D)^{bd}-\frac{1}{g\ell^{2}}\epsilon^{abc}(\partial_{\mu}f^{b}\cdot
D^{cd}_{\mu}).
\end{equation}
Since we will do a background decomposition as given by equation
(\ref{eq:2}) to find the vacuum to vacuum functional in the
presence of $\tilde{f}^{a}(x)$ and then do a stochastic averaging,
we will change the delta functionals by
\begin{eqnarray}\label{eq:28}
\delta(\partial\cdot
t^{a}-\frac{1}{g\ell^{2}}f^{a})&=&det(\frac{1}{(1+\vec{f}\cdot\vec{f})^{j}})
\delta(\frac{1}{(1+\vec{f}\cdot\vec{f})^{j}}(\partial\cdot
t^{a}-\frac{1}{g\ell^{2}}f^{a}))\nonumber\\
\delta(\rho^{a})&=&det(\frac{1}{(1+\vec{f}\cdot\vec{f})^{k}})\delta(\frac{1}{(1+\vec{f}\cdot\vec{f})^{k}}\rho^{a})
\end{eqnarray}
where the powers $j$ and $k$ can be freely chosen so that
stochastic averages related to the gauge-fixing and the
"Fadeev-Popov" determinant vanish.

To see how this happens, let us write the path-integral as
\begin{equation}\label{eq:29}
W=\int(dt^{a}_{\mu})(df^{a})(d\psi)(d\bar{\psi})(du^{a})(d\bar{u}^{a})exp.\{-S'\},
\end{equation}
where
\begin{eqnarray}
S'&=&\int d^{4}x \mathcal{L}'(x),\nonumber\\
\mathcal{L}'(x)&=&\mathcal{L}_{YM}+\mathcal{L}_{fermion}+\mathcal{L}_{gf}+\mathcal{L}_{FP}\label{eq:30}\\
\mathcal{L}_{gf}&=&(\frac{1}{\alpha})\frac{1}{(1+\vec{f}\cdot\vec{f})^{2j}}(\partial\cdot
t^{a}-\frac{1}{g\ell^{2}}f^{a})^{2}+(\frac{1}{\beta})\frac{1}{(1+\vec{f}\cdot\vec{f})^{2k}}\rho^{a}\rho^{a}\label{eq:31}\\
\mathcal{L}_{FP}&=&\bar{u}^{a}\frac{1}{[(1+\vec{f}\cdot\vec{f})^{j+k+4}]}\theta^{ab}u^{b}\label{eq:32}.
\end{eqnarray}
The ghosts $\bar{u}^{a}$, $u^{a}$ are introduced to express the
determinants.  Next, we introduce the background decomposition
given by equation (\ref{eq:2}) and equation (\ref{eq:29}) will
yield
\begin{equation}\label{eq:33}
W[\tilde{f}^{a}]=\int(dt^{a}_{\mu})(d\psi)(d\bar{\psi})(d\phi^{a})(d\bar{u}^{a})(du^{a})exp.
\{-S'(t,\phi,\psi,\bar{\psi},u^{a},\bar{u}^{a};\tilde{f})\}
\end{equation}

Next we do the stochastic averaging given by
\begin{equation}\label{eq:34}
\langle
W[\tilde{f}^{a}]\rangle_{\tilde{f}^{a}}=\int(dt^{a}_{\mu})(d\psi)(d\bar{\psi})(d\phi^{a})(du^{a})(d\bar{u}^{a})\langle
exp.\{-S'\}\rangle_{\tilde{f}}
\end{equation}
We evaluate the stochastic average by expanding $e^{-s'}$. The
following averages are needed:
\begin{eqnarray}
\langle S'\rangle = \langle
S_{YM}+S_{fermion}+S_{gf}+S_{FP}\rangle_{\tilde{f}},\label{eq:35}\\
\langle S'^{2}\rangle =\langle S'\rangle^{2}+\int d^{4}x
d^{4}y\langle
\underbrace{\mathcal{L}'(x)\mathcal{L}'(y)}\rangle_{\tilde{f}},\label{eq:36}\\
\langle S'^{3}\rangle =\langle S'\rangle^{3}+3(\int d^{4}x
d^{4}y\langle
\underbrace{\mathcal{L}'(x)\mathcal{L}'(y)}\rangle_{\tilde{f}})\langle S' \rangle,\label{eq:37}\\
\langle S'^{4}\rangle =\langle S'\rangle^{4}+3(\int d^{4}x
d^{4}y\langle
\underbrace{\mathcal{L}'(x)\mathcal{L}'(y)}\rangle_{\tilde{f}})\langle S' \rangle^{2},\nonumber\\
+\int d^{4}x d^{4}y d^{4}z d^{4}r\langle
\underbrace{\mathcal{L}'(x)\mathcal{L}'(y)\mathcal{L}'(z)\mathcal{L}'(r)}
\rangle_{\tilde{f}}.\label{eq:38}
\end{eqnarray}

In equation (\ref{eq:35}), we make use of equations (\ref{eq:18})
and (\ref{eq:19}). In equations (\ref{eq:36},\ref{eq:37}) and
(\ref{eq:38}), the symbols
$\underbrace{\mathcal{L}'(x)\mathcal{L}'(y)}$, etc., represent the
correlated points, which arise because the derivative of the
white-noise $\tilde {f}^{a}(x)$ is "smoothed out" via
\begin{equation}\label{eq:39}
\frac{d\tilde{f}^{a}}{dx}=\frac{\tilde{f}^{a}(x+\frac{\ell}{n})-\tilde{f}^{a}(x)}{\frac{\ell}{n}}
\end{equation}
Note that all odd correlations, such as
$\langle\underbrace{\mathcal{L}'(x)\mathcal{L}'(y)\mathcal{L}'(z)}\rangle_{\tilde{f}}$
vanish because of the white-noise character of $\tilde{f}^{a}(x)$.
The stochastic averages involve the following integral
\begin{equation}\label{eq:40}
\lim_{\sigma\rightarrow
0}\pi^{-3/2}\sigma^{+3/2}\int^{\infty}_{0}\frac{r^{2m}}{(1+r^{2})^{n}}e^{-\sigma
r^{2}}dr = \left\{\begin{array}
    {r@{\quad\quad}l}
    0,& \mbox{for}\,\,m\leq n\\ & \\

    finite,&  \mbox{for}\,m=n+1\\ & \\

    diverges,&  \mbox{for}\,m \geq 0, n+2.
    \end{array}\right.
\end{equation}

Using equation (\ref{eq:40}) and because of the
$(1+\tilde{f}\cdot\tilde{f})^{2}$ factors in the denominators of
$S_{gf}$ and $S_{FP}$, we find that when we do a background
decomposition given by equation (\ref{eq:2}), the stochastic
averages of each of the expansion terms vanish.  Thus, we find
\begin{equation}\label{eq:41}
\langle S_{gf}\rangle_{\tilde{f}}=\langle
S_{FP}\rangle_{\tilde{f}}=0,
\end{equation}
giving $\langle S'\rangle = \langle S\rangle$, which is given by
equations (\ref{eq:18}) and (\ref{eq:19}).  Furthermore, the
correlated terms in equations (\ref{eq:36}), (\ref{eq:37}) and
(\ref{eq:38}) that involve $\mathcal{L}_{gf}$ and
$\mathcal{L}_{FP}$ also vanish for the same reason.

Taking everything so far into account, we find
\begin{eqnarray}\label{eq:42}
\langle e^{-S'}\rangle_{\tilde{f}}&=& 1-\langle S\rangle+\frac{1}{2}[\langle S\rangle^{2}\nonumber\\
&+&\int
d^{4}xd^{4}y\langle\underbrace{\mathcal{L}(x)\mathcal{L}(y)}\rangle_{\tilde{f}}]
-\frac{1}{3!}[\langle S\rangle^{3}\nonumber\\
&+&3[(\int
d^{4}xd^{4}y)(\langle\underbrace{\mathcal{L}(x)\mathcal{L}(y)}\rangle_{\tilde{f}})\langle
S\rangle]\nonumber\\
&+&\frac{1}{4!}[\langle S\rangle^{4}+6(\int
d^{4}xd^{4}y\langle\underbrace{\mathcal{L}(x)\mathcal{L}(y)}\rangle_{\tilde{f}})\langle
S\rangle^{2}\nonumber\\
&+&\int d^{4}xd^{4}yd^{4}zd^{4}r\langle
\underbrace{\mathcal{L}(x)\mathcal{L}(y)\mathcal{L}(z)\mathcal{L}(r)}\rangle_{\tilde{f}}]\nonumber\\
&+&....
\end{eqnarray}

Summing the series gives
\begin{equation}\label{eq:43}
\langle e^{-S'}\rangle_{\tilde{f}}=e^{-S_{eff}},
\end{equation}
where
\begin{eqnarray}\label{eq:44}
S_{eff}&=&\langle S\rangle -\frac{1}{2}\int d^{4}xd^{4}y\langle
\underbrace{\mathcal{L}(x)\mathcal{L}(y)}\rangle_{\tilde{f}}\nonumber\\
&-&\frac{1}{4!}\int
d^{4}xd^{4}yd^{4}zd^{4}r\langle\underbrace{\mathcal{L}(x)\mathcal{L}(y)\mathcal{L}(z)\mathcal{L}(r)}\rangle_{\tilde{f}}\nonumber\\
&+&....
\end{eqnarray}

Since equation (\ref{eq:44}) is ghost independent, we simply drop
the ghost path-integral and lump it with the normalization factors
needed to make sense of the path-integral.

At this point, where is the non-local four-fermi (NLFF)
interaction in equation (\ref{eq:44})?  It is found in
\begin{equation}\label{eq:45}
NLFF=\frac{g^{2}}{2}\int
d^{4}xd^{4}y(\bar{\psi}\gamma_{\mu}T^{a}\psi)_{x}\langle
A^{a}_{\mu}(x)A^{b}_{\nu}(y)\rangle_{\tilde{f}}(\bar{\psi}\gamma_{\nu}T^{b}\psi)_{y}.
\end{equation}
We will evaluate the stochastic average by making use of
\begin{equation}\label{eq:46}
\partial_{\mu}A^{a}_{\mu}(x)=\frac{1}{g\ell^{2}}f^{a}(x)=\frac{1}{g\ell^{2}}\tilde{f}^{a}(x)+\frac{1}{g\ell^{2}}\phi^{a}(x).
\end{equation}

From equation (\ref{eq:46}), we must have
\begin{equation}\label{eq:47}
\partial^{x}_{\mu}\partial^{y}_{\nu}\langle
A^{a}_{\mu}(x)A^{b}_{\nu}(y)\rangle_{\tilde{f}}=\frac{1}{g^{2}\ell^{3}}\delta(x-y)+...,
\end{equation}
where $x=(x_{\mu}x_{\mu})^{1/2}$ and $y=(y_{\mu}y_{\mu})^{1/2}$.
Equation (\ref{eq:47}) implies that
\begin{equation}\label{eq:48}
\langle
A^{a}_{\mu}(x)A^{b}_{\nu}(y)\rangle=(\frac{1}{g^{2}\ell^{3}})\frac{x_{\mu}}{x}\frac{y_{\nu}}{y}\delta^{ab}|x-y|+...
\end{equation}

The equivalence follows from the fact that for a spherically
symmetric function,
$\partial^{x}_{\mu}=\frac{x_{\mu}}{x}\frac{d}{dx}$ and
$\frac{d^{2}}{dx^{2}}|x-y|=\delta(x-y)$.  Substituting equation
(\ref{eq:47}) in NLFF, we find
\begin{equation}\label{eq:49}
NLFF=\frac{1}{2}(\frac{1}{\ell^{3}})\int
d^{4}xd^{4}y(\bar{\psi}\eta_{\mu}\gamma_{\mu}
T^{a}\psi)_{x}|x-y|(\bar{\psi}\gamma_{\nu}\eta_{\nu}T^{a}\psi)_{y}+...
\end{equation}
where
\begin{eqnarray}\label{eq:50}
\eta_{\mu}=(sin\theta_{1} sin\theta_{2} sin\phi, sin\theta_{1}
sin\theta_{2} cos\phi, sin\theta_{1} cos\theta_{2},cos\theta_{1}),
\end{eqnarray}
i.e., $\eta_{\mu}$ represents the unit vectors in 4D spherical
coordinates.  If the fermion field is spherically symmetric, the
angular integration does not vanish only when
$\eta_{\mu}(\vec{x})=\pm\eta_{\mu}(\vec{y})$, i.e., the 4D vectors
are collinear.  Using

\begin{equation}
\int d
\Omega_{4}\eta_{\mu}\eta_{\nu}=\frac{\pi^{2}}{2}\delta_{\mu\nu}\label{eq:51}
\end{equation}
and $\int
d\Omega_{4}=\int^{2\pi}_{0}\int^{\pi}_{0}\int^{\pi}\sin^{2}\theta_{1}
\sin\theta_{2}d\theta_{1}d\theta_{2}d\phi=2\pi^{2}$, we find that
we can write the equation (\ref{eq:49}) as
\begin{equation}\label{eq:52}
NLFF=\frac{1}{8}\frac{1}{\ell^{3}}\int
d^{4}xd^{4}y(\bar{\psi}\gamma_{\mu}\psi)_{x}|\vec{x}-\vec{y}|(\bar{\psi}\gamma_{\mu}\psi)_{y}
\end{equation}
with $\vec{x}$ and $\vec{y}$ collinear This hints of flux tube
geometry as shown in figure~\ref{fig:1}:
\begin{figure}
\centering
\includegraphics[totalheight=2.5 in]{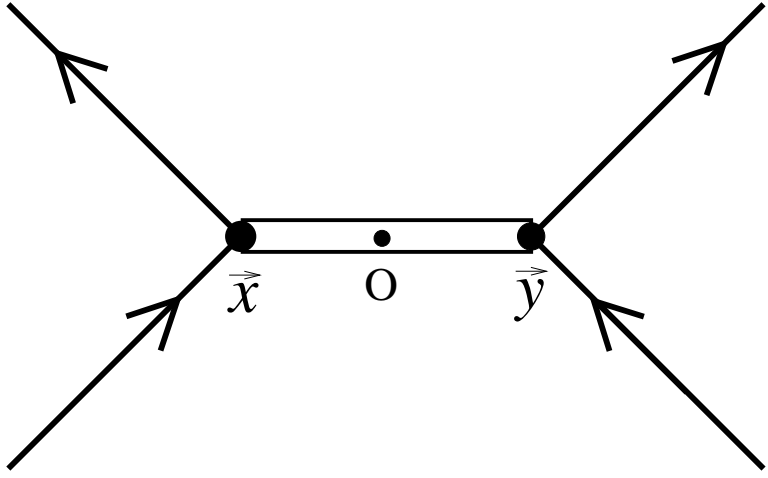}
\caption{The four-fermi term with interaction constrained with
$\vec{x}$,$\vec{y}$ collinear hinting of a flux tube geometry}
\label{fig:1}
\end{figure}

This result may be a bit surprising because we started with an
expansion about a classical background $\tilde{f}^{a}(x)$, which
is spherically symmetric.  But considering that the r potential is
no less or no more spherically symmetric than the $\frac{1}{r}$
term, this result is not unreasonable at all.

When we determined the correlation $\langle
A^{a}_{\mu}(x)A^{b}_{\nu}(y)\rangle_{\tilde{f}}$ to arrive at the
linear term (note there are extra terms), we made use of
components of the vector field involving $\phi^{a},t^{a}_{\mu}$ in
a background $\tilde{f}^{a}(x)$.  All these components are
important for the following reasons: (1) As was shown in Section
II, just considering $t^{a}_{\mu}$ alone after stochastic
averaging, does not lead to a confining term.  (2) As for the
$\phi^{a}$ alone, it has no kinetic term and if we integrate it
out yields a constraint on the fermion fields, which under the
assumption of spherical symmetry says the fermion field must
vanish beyond $\ell$ (see reference (18)). (3) Considering
$\tilde{A}^{a}_{\mu}({\tilde{f}})$ alone will not do either
because $\partial_{\mu}\tilde{A}^{a}_{\mu}(\tilde{f})=0$, if we
neglect $\phi^{a}$ and $t^{a}_{\mu}$, and we needed the
$\tilde{f}^{a}_{\mu}(x)$ at the right-hand side of equation
(\ref{eq:46}) to derive equation (\ref{eq:48}).  Thus, a certain
combination of $\phi^{a},t^{a}_{\mu}$ in $\tilde{f}^{a}(x)$ is key
in deriving the confining non-local four-fermi term. Determining
which components these are is difficult from a direct evaluation
because there are many terms that contribute to it. The linear
term appears but an unusual"renormalization", to get rid of a
$\frac{1}{\sigma}$ term with $\sigma\longrightarrow 0$, must be
carried out.  Thus, the derivation made use of equation
(\ref{eq:46}) where, the divergence does not show up.

\section{Conclusion}

In this paper, we showed confinement even with dynamical quarks.
The confinement mechanism is the stochastic treatment of the
scalar classical configurations $\tilde{f}^{a}(x)$, which arise in
the non-linear regime of the non-linear gauge.

\section{Acknowledgement}

This research was supported in part by the Natural Sciences
Research Institute of the University of the Philippines.



\end{document}